\documentclass[twocolumn,showpacs,preprintnumbers,amsmath,amssymb]{revtex4}

\usepackage{epsfig,psfrag}
\usepackage{dcolumn}
\usepackage{bm}
\usepackage{graphicx}
\usepackage{color}

\begin{document}


\title{Josephson current through a nanoscale magnetic quantum dot}

\author{F. Siano and R. Egger}

\affiliation{Institut f\"ur Theoretische Physik, 
Heinrich-Heine-Universit\"at,
 D-40225 D\"usseldorf, Germany}

\date{\today}

\begin{abstract}
We present theoretical results 
for the equilibrium Josephson current through
an Anderson dot tuned into the magnetic regime, using
Hirsch-Fye Monte Carlo simulations covering
the complete crossover from Kondo-dominated physics
to $\pi$ junction behavior in a numerically exact way.
Within the `magnetic' regime, 
$U/\Gamma\gg 1$ and  $\epsilon_0/\Gamma\leq 1$,
the Josephson current is found to
depend only on $\Delta/T_K$, where $\Delta$ is the   
BCS gap and $T_K$ the Kondo temperature.  The junction behavior can
be classified into four different quantum phases.
We describe these behaviors, specify the associated
three transition points, and identify a local minimum in the
 critical current of the junction as a function of $\Delta/T_K$.
\end{abstract}


\maketitle

\begin{figure} 
\scalebox{0.5}{ \includegraphics{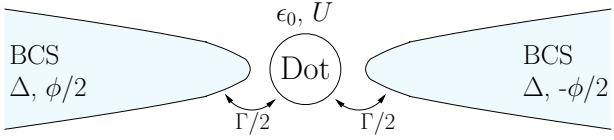} }
\caption{\label{fig0} 
Schematic setup of an Anderson dot 
with local energy $\epsilon_0$ and charging energy $U>0$  
coupled to BCS leads. For simplicity, we assume identical BCS gap $\Delta$
on both sides, with phase difference $\phi$ across
the dot and hybridization $\Gamma$.  }
\end{figure}

Recent advances in nanoscale manipulation and fabrication 
call for a deeper understanding of the
effect of electronic correlations.  Due to the 
complexity of its theoretical treatment,  
the interplay between superconductivity and magnetism 
belongs to the least understood phenomena in that respect.
Here we study the Josephson current 
$I(\phi)$ through a correlated nanoscale quantum dot contacted by $s$-wave
BCS superconductors. 
At low enough temperatures, such a dot
is generally described by the Anderson impurity model 
indicated in Fig.~\ref{fig0}.
We consider the regime
 $U/\Gamma\gg 1$ and $\epsilon_0/\Gamma\ll  -1$, where the dot  effectively has
single occupancy and thus represents a spin-1/2 degree
of freedom.  
Then a complicated interplay between this magnetic impurity and the 
superconductivity in the leads sets in.  Some aspects of this
physics were recently observed in Andreev conductance measurements for a 
short multi-wall nanotube \cite{buitelaar1,buitelaar2}.  A similar setup
should also allow to probe the Josephson current in the near
future, where  the ratio $\Delta/T_K$ is widely tunable via a backgate voltage.

In this paper, we provide a detailed 
analysis and classification of all
possible phases expected in such an experiment.
We find that only one `master' parameter  
$\Delta/T_K$ governs this problem, where 
\begin{equation}\label{tk}
T_K =  \frac12\sqrt{\Gamma U} \exp[\pi \epsilon_0 (\epsilon_0+U)/\Gamma U]
\end{equation}
is the Kondo temperature for normal leads \cite{haldane}. 
For $\Delta/T_K\ll 1$, the Kondo effect survives
and is only weakly affected by superconductivity,
while for $\Delta/T_K\gg 1$,
perturbation theory in $\Gamma$ yields an inverted Josephson 
relation $I(\phi)=-I_c \sin\phi$ \cite{kulik,shiba,glazman,spivak},
where  $\phi=\pi$ represents a minimum of the 
junction free energy $F(\phi)$.
Such a $\pi$ junction behavior
was recently reported in Nb-Cu$_x$Ni$_{1-x}$-Nb systems \cite{ryan},
is related to subgap (Andreev) bound states \cite{arovas,clerk},
and implies broken time-reversal symmetry.
In both limits, analytical expressions \cite{glazman}
are reproduced by our method below.
For a magnetic impurity, 
the Josephson relation is generally replaced by a more 
complicated dependence on $\phi$.
A classification into four types of junctions, labeled as
{\boldmath $0$}, {\boldmath $0^\prime$},  {\boldmath $\pi^\prime$}
and {\boldmath $\pi$}, follows from the respective stability
of the $\phi=0$ and $\phi=\pi$ configurations \cite{arovas}.
For a {\boldmath $0$} ({\boldmath $\pi$}) junction,
only $\phi=0 \ (\phi=\pi$) is a 
minimum of $F(\phi)$.  For the two other cases,
both $\phi=0,\pi$ are {\sl local}\
minima, and depending on whether
 $\phi=0 \ (\phi=\pi)$ is the {\sl global}\ minimum, 
one has a {\boldmath $0^\prime$}  ({\boldmath $\pi^\prime$})
junction. 
Using $I(\phi)=(2e/\hbar) dF(\phi)/d\phi$, the phase boundaries can
be directly read off from the $I(\phi)$ curves. For instance, 
the {\boldmath $0'$}-{\boldmath $\pi'$} transition point is determined by the
condition $\int_0^\pi d\phi I(\phi)=0$.
 
The theoretical description of the
resulting phase diagram is difficult,
and despite intense efforts over the past few years
\cite{arovas,clerk,ando,yosh,roz,guinea,mats,kozub,vecino,belzig},
a satisfactory physical picture has not been obtained so far.
This paper provides numerically exact results for a
magnetic Josephson junction from 
Monte Carlo (MC) simulations, using
a Hirsch-Fye algorithm \cite{hirsch,fye,kusa} adapted to this
problem.  We find that previous approximate theories for this
problem, relying on
the non-crossing approximation (NCA) for $U\to \infty$
\cite{clerk,ando,roz}, mean-field approaches \cite{arovas,vecino},
perturbative schemes \cite{guinea,mats,kozub,vecino}, or the
numerical renormalization group (NRG) \cite{yosh,belzig},
lead to incomplete and sometimes even qualitatively inaccurate predictions.  
Although the  existence
of the above-mentioned phases follows
already from mean-field theory \cite{arovas}, their respective stability and
the actual phase boundaries have not been reliably established. 
Moreover, the critical current  defined by
\begin{equation}
\label{ic}
I_c (\Delta/T_K) =  {\rm max}_{\phi} \ | I(\phi,\Delta/T_K) | 
\end{equation}
has a {\sl non-monotonic behavior}
as a function of $\Delta/T_K$ which has been missed by
all previous studies.   The predicted minimum as well as 
the phase boundaries specified in 
Eqs.~(\ref{firsttrans},\ref{secondtrans},\ref{thirdtrans}) below
should be observable in state-of-the-art experiments.

We study the `canonical' model for this problem, see
Refs.~\cite{glazman,arovas,clerk,ando,yosh,roz,guinea,mats,kozub,vecino,belzig}
and Fig.~\ref{fig0}.  
For a symmetric situation,  
\begin{eqnarray} \label{ham}
H &=&  \sum_{\vec k,p}  \Bigl[
\epsilon_k \sum_\sigma c_{\sigma,p}^\dagger (\vec k)
c_{\sigma,p}^{} (\vec k) \\  \nonumber
&-& \left (\Delta e^{ip \phi/2} c_{\uparrow,p}^\dagger(\vec k)
c_{\downarrow,p}^\dagger(-\vec k) + {\rm h.c.} \right)  \\ \nonumber
&-& \sum_\sigma \left(
 t c_{\sigma,p}(\vec k) d^\dagger_\sigma + {\rm h.c.}\right)\Bigr]
\\ \nonumber
&+&
\left(\epsilon_0+\frac{U}{2}\right) (n_\uparrow+n_\downarrow) - 
\frac{U}{2} (n_\uparrow
-n_\downarrow)^2,
\end{eqnarray}
where $c_{\sigma,p}(\vec k)$ is the electron operator for lead $p=L/R=\pm$
and spin $\sigma=\uparrow,\downarrow$, with single-particle dispersion 
$\epsilon_k$.  Moreover, $n_\sigma= d^\dagger_\sigma d^{}_\sigma$ with the
dot electron operator $d_\sigma$, and
$\Gamma= 2\pi \rho_0 |t|^2$ for lead density of states $\rho_0$
and hopping matrix element $t$.
In all simulations reported below, $T/\Delta=0.1$ (we put $\hbar=k_B=1$),
which corresponds to a temperature $T\approx 100$~mK for the setup of 
Refs.~\cite{buitelaar1,buitelaar2}.
By comparing to analytical results for  
$\Delta/T_K \gg 1$ and $\ll 1$, this appears to be quite close to the
ground-state limit.  

To formulate the MC scheme, we construct the
imaginary-time path integral representation under this model for the 
Josephson current.
Discretizing imaginary time in steps of size $\delta=1/PT$ for $P$
discretization points, 
the last term in Eq.~(\ref{ham})
can be decoupled by auxiliary Ising spins $s_k=\pm 1$ defined at 
times $\tau_k =k\delta$, where $k=1,\ldots,P$, using the
discrete Hubbard-Stratonovich transformation \cite{hirsch}
\begin{equation}
e^{ (U\delta/2) (n_\uparrow-n_\downarrow)^2(\tau_k) }=
 \frac12 \sum_{s_k=\pm 1} e^{-\lambda s_k (n_\uparrow
- n_\downarrow)(\tau_k)},
\end{equation}
where $\lambda = \cosh^{-1}[\exp(U\delta/2)]$.
All fermions characterizing the dot and the leads are
then free and can be integrated out. 
Thereby the Josephson current is expressed 
in terms of a cyclic 1D Ising spin chain 
with non-standard long-ranged spin-spin interactions,
\begin{equation}\label{ij}
\frac{I(\phi)}{I_0}  =  \frac{ 
 \sum_{\{ s \} }  {\rm det}( \hat{G}^{-1}+ \lambda \hat{s }) 
 \ {\rm Tr}\left\{ [\hat{G}^{-1}+\lambda \hat{s}]^{-1} \hat{\Sigma}^J \right\} }
{\sum_{\{ s \} }  {\rm det}( \hat{G}^{-1}+ \lambda \hat{s}  ) }, 
\end{equation}
where $I_0=e\Delta/\hbar$ is the critical current in the unitary
limit, $\hat{s}$ has matrix elements
$\hat{s}_{ij, \mu\nu}= s_i \delta_{ij}
\delta_{\mu\nu}$ with Nambu indices $\mu,\nu=1,2$ and time
indices $i,j$, and
\begin{eqnarray}
\nonumber
\hat{G}^{-1}_{ij} &=& \delta^2 T \sum_{\omega_n} e^{-i\omega_n
\delta (i-j) } \Bigl [-i \omega_n \tau_0 + (\epsilon_0+U/2) \tau_3 \\
\label{gij}
&-&  \frac{\Gamma}{\sqrt{\omega_n^2+\Delta^2}}
 \left(i\omega_n \tau_0 + \Delta\cos(\phi/2) \tau_1\right)  \Bigr ].
\end{eqnarray}
The Josephson self energy  is
\begin{equation}
\hat{\Sigma}^J_{ij} = - (\delta T)^2 \Gamma \sin(\phi/2) 
  \tau_1 \sum_{\omega_n}
 \frac{e^{-i\omega_n \delta (i-j)}
}{\sqrt{\omega_n^2+\Delta^2}}.
\end{equation}
All summations over fermion Matsubara
 frequencies $\omega_n=(2n+1)\pi T$ are 
 restricted to $|\omega_n|<\pi/\delta$,
since finite $\delta$ implies the existence of a UV cutoff.
The Pauli matrices $\tau_i$ act in Nambu space, 
with $\tau_0={\rm diag}(1,1)$. We mention that
a modified version of this approach could also access the case of
unconventional superconductor leads, where additional
features are expected \cite{newref,aono}.

Equation (\ref{ij}) allows to compute the equilibrium Josephson current
for arbitrary parameters under a MC scheme.
Finite-$\delta$ corrections can be eliminated by exploiting the theorem
that such corrections must scale $\propto \delta^2$ for any
Hermitian observable \cite{fye}. For given physical parameters,
we thus compute $I(\phi)$
for a few (small) values of $\delta$, and 
then perform the extrapolation
$\delta\to 0$ using a linear-regression fit.
The $\delta^2$ scaling is well obeyed for $\Delta\delta \leq 0.1$
even for $U/\Delta=20,\Gamma/\Delta=5$,
 leading to discretization numbers  $P\approx$ 140 to 270.
Then no systematic errors are present, and numerical
results are exact within stochastic MC error bars.
Due to the absence of particle-hole symmetry,
this algorithm has a sign problem  for $\Delta\ne 0$,
which manifests itself in occasionally negative
determinants in Eq.~(\ref{ij}). Fortunately,  this problem
is very weak in our implementation, 
with average sign above $0.7$,  
and hence does not restrict the method in practice.  
Local flip updates of the Ising spins under the standard Metropolis 
algorithm were sufficient to ensure rapid equilibration
and satisfactory MC acceptance rates.
Under a spin flip $s_k\to -s_k$, 
corresponding to 
$\lambda \hat {s} \to\lambda
 \hat {s} + \hat{ w}$, where $w_{ij,\mu\nu}=-2\lambda
s_k \delta_{ik}\delta_{jk} \delta_{\mu\nu}$,
the change in weight 
is given by the ratio $R$ of new and old determinants appearing
in Eq.~(\ref{ij}), which can be evaluated analytically. With the matrix 
$\hat{D}= (\hat{G}^{-1}+ \lambda \hat{s})^{-1}$,  we find
\begin{eqnarray*}
R & = & {\rm det}(1+\hat {D} \hat{w}) = 
 1 -2 \lambda s_k (  D_{kk,11}+D_{kk,22} ) \\
&&+ 4 \lambda^2  
[ D_{kk,11} D_{kk,22} - D_{kk,12} D_{kk,21} ].
\end{eqnarray*}
Thereby the costly explicit calculation of determinants is avoided.
Similarly, $\hat{D}$ can be updated without explicit matrix inversion.
Typically,  $10^6$ MC samples were accumulated to obtain
each data point below. On a 2 GHz Xeon processor, our code performs at
a speed of 3.7 CPU hours per $10^5$ samples for $P=180$.  

\begin{figure} 
\scalebox{0.35}{
\rotatebox{270}{
\includegraphics{{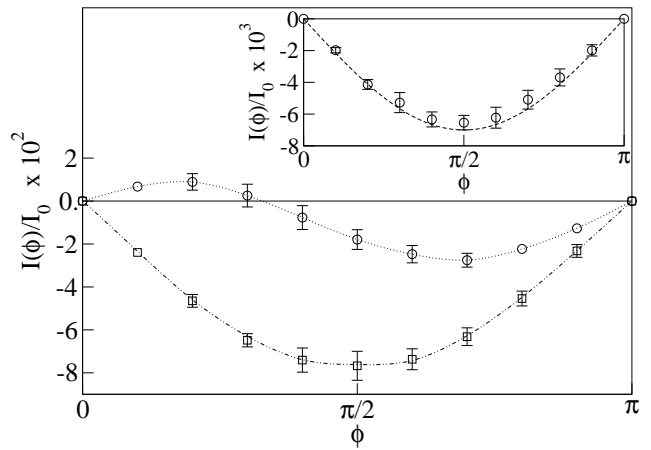}}
}}
\caption{\label{fig1} 
MC results for the 
Josephson current in units of $I_0=e\Delta/\hbar$ 
for two parameter sets with $\Delta/T_K\approx 23$.
Squares are for $U/\Delta=20, \Gamma/\Delta=3.44$, 
and circles for $U/\Delta=4,\Gamma/\Delta=1$.
Unless noted otherwise, in all figures, curves are guides to the eye only.  
Stochastic MC errors are always smaller than the symbol size or indicated by
vertical bars.
Inset: Current-phase relation in the deep $\pi$ junction regime.
MC results for $\Delta/T_K=6.5\times 10^4$ coincide with
the analytical result (dashed curve) \cite{glazman}.
}
\end{figure}

First, the code was checked against analytical solutions
 available for small and large $\Delta/T_K$ \cite{glazman},
which were accurately reproduced, see, for instance, the
inset of Fig.~\ref{fig1} for $\Delta/T_K\gg 1$.
Our simulations are therefore able
to cover the complete crossover from Kondo-dominated physics
to a $\pi$ junction.  (The case $\Delta=0$ has also been studied
in Refs.~\cite{hirsch,fye} using this method.) 
Moreover, we have checked for many parameter sets that 
universality is fulfilled, 
i.e., taking different values for $\epsilon_0/\Delta, 
U/\Delta$ and/or $\Gamma/\Delta$  only affects the Josephson current
via the corresponding change in $\Delta/T_K$.
However,  the two curves for $\epsilon_0=-U/2$ shown
 in the main part of Fig.~\ref{fig1} demonstrate
that profound differences can arise even 
for the same $\Delta/T_K$, if $U/\Gamma$ is not sufficiently large. 
We found $U/\Gamma\geq 5$ necessary to
 ensure universality; otherwise
charge fluctuations may alter the Josephson
current even qualitatively, see Fig.~\ref{fig1}.
Universality then holds only in the true magnetic regime, and not
away from it.  We mention in passing that 
the critical current never exceeded $I_0$, in contrast to 
the prediction in Ref.~\cite{mats}.

\begin{figure} 
\scalebox{0.35}{
\rotatebox{270}{
\includegraphics{{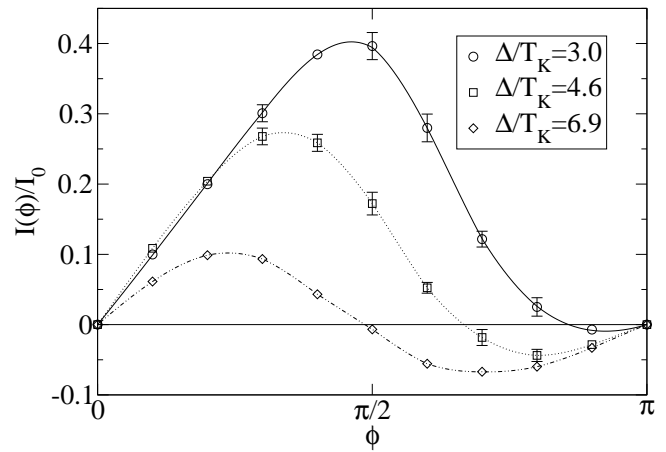}}
}}
\caption{\label{fig2} 
MC data for the current-phase relation at small-to-intermediate
 $\Delta/T_K$. 
} 
\end{figure}

\begin{figure} 
\scalebox{0.35}{
\rotatebox{270}{
\includegraphics{{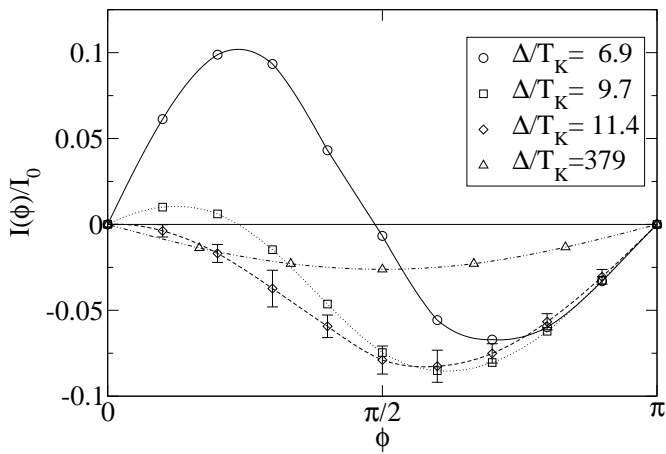}}
}}
\caption{\label{fig3} Same as Fig.~\ref{fig2}
for intermediate-to-large $\Delta/T_K$.  }
\end{figure}

\begin{figure} 
\scalebox{0.35}{
\rotatebox{270}{
\includegraphics{{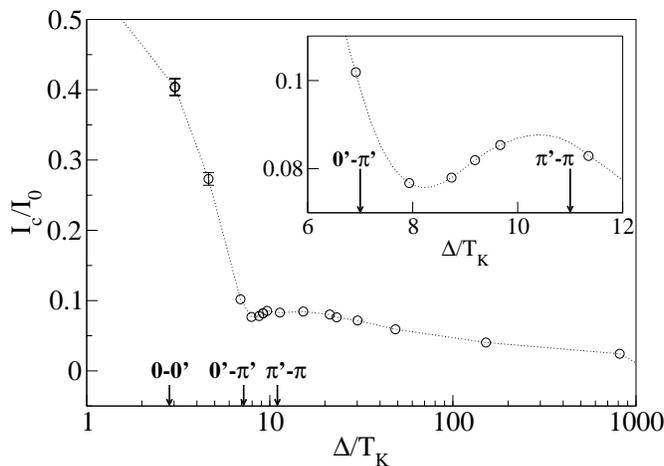}}
}}
\caption{\label{fig4} MC results for the critical current (\ref{ic}) 
as a function of $\Delta/T_K$.
The arrows mark the boundaries between different phases (see text).
The inset shows the region around the minimum on a magnified scale.
Error bars are always smaller than symbol size.
}
\end{figure}

We then continue by discussing the full crossover between these
two limiting cases. Numerical results shown below were  
obtained for $5\leq U/\Gamma \leq 10$ 
and $0.1\leq \Delta/\Gamma \leq 10$,  putting $\epsilon_0=-U/2$.
In Figures \ref{fig2} and \ref{fig3}, representative 
 data for $I(\phi)$ covering this crossover are presented.  
Furthermore, in Fig.~\ref{fig4} we show the critical current 
(\ref{ic}) as a function of $\Delta/T_K$.
For very small $\Delta/T_K$, the Kondo effect is dominant, and we find a 
 {\boldmath $0$} junction with a non-sinusoidal current-phase
relation close to the unitary limit, $I_c=I_0$ \cite{glazman}.
For larger  $\Delta/T_K$,  the 
relation is more sinusoidal again,  and
with increasing $\Delta/T_K$, the critical current $I_c$
decreases. Moreover, 
above a first transition point 
\begin{equation} \label{firsttrans}
(\Delta/T_K)_{00'} = 2.8 \pm 0.1,
\end{equation}
for $\phi$ slightly below $\pi$, the current is negative, and
under the above classification scheme, we thus 
enter the {\boldmath $0'$} phase, see Fig.~\ref{fig2}.
By increasing $\Delta/T_K$ further, one eventually 
reaches a second transition point at 
\begin{equation}\label{secondtrans}
(\Delta/T_K)_{0'\pi'}= 7.2 \pm 0.2.
\end{equation}
At the transition point,  $\int d\phi \
I(\phi) = 0$.  We have now reached the {\boldmath $\pi'$} phase, 
where $\phi=\pi$ is already the global minimum, but $\phi=0$ still
represents a local minimum. The true
 {\boldmath $\pi$} phase, with $\phi=\pi$ as the
 only minimum, is eventually reached at
\begin{equation}\label{thirdtrans}
(\Delta/T_K)_{\pi'\pi}=11.0 \pm 0.3.
\end{equation}
For $\Delta/T_K\gg (\Delta/T_K)_{\pi'\pi}$, the inverted
Josephson relation $I=-I_c\sin\phi$ valid in the
deep $\pi$ junction limit \cite{glazman} is finally recovered, see
Figs.~\ref{fig1} and \ref{fig3}.

The transition points reported here are at quite different locations
than thought previously.  For instance, NRG calculations \cite{yosh,belzig} 
find that the transition into the {\boldmath $\pi$} phase
occurs at $\Delta/T_K\approx 2.4$, where according to our data
the junction is still in the {\boldmath $0$} phase.   
The differences to NCA and/or mean-field results are even more drastic.
According to our simulation data,
the {\boldmath $\pi$} phase covers a much smaller
region in parameter space than thought previously, while the 
intermediate {\boldmath $0'$} and {\boldmath $\pi'$} phases 
extend over a significant range in $\Delta/T_K$, and therefore should
be readily observed in practice.  
Remarkably, the critical current shown in Fig.~\ref{fig4} displays
non-monotonic behavior. {}From the analytically known
limits, $I_c(\Delta/T_K\to 0)\to I_0$ and $I_c(\Delta/T_K\to \infty)\to
0$, a naive guess is to expect that $I_c(\Delta/T_K)$ just
drops monotonically with increasing $\Delta/T_K$. Such 
a behavior is in fact predicted by 
previous work, see, e.g., Ref.~\cite{belzig}.  
Our simulations point to a more complicated picture,
  where $I_c(\Delta/T_K)$ is characterized by a local {\sl minimum}.
This minimum occurs at
\begin{equation}
\label{minimum}
(\Delta/T_K)_{\rm min} = 8.2\pm 0.2,
\end{equation} 
which is close to but above the transition point (\ref{secondtrans})
for the {\boldmath $0'$}-{\boldmath $\pi'$} transition.
After reaching a local maximum slightly below the 
{\boldmath $\pi'$}-{\boldmath $\pi$} transition, the
critical current then drops monotonically throughout the {\boldmath $\pi$}
regime.  At first sight, the local minimum in $I_c$ appears to resemble the 
experimentally observed  oscillations in the critical current as a 
function of either temperature or length of the junction \cite{ryan}.  
However, while those oscillations can be traced back to
Andreev bound state crossings, such a simple, essentially mean-field type
reasoning does not apply here, see also Ref.~\cite{vecino} for a 
closely related discussion. 
The appearance of a local minimum in $I_c$ thus indicates a 
subtle many-body effect.

To conclude, we have presented a numerically exact
study of the Josephson current through a nanoscale dot. 
In the magnetic regime, $U/\Gamma\gg 1$ and $\epsilon_0/\Gamma\leq 1$,
our results reveal a rather complex behavior that is 
governed by $\Delta/T_K$ as the only tuning parameter.
These predictions should be observable in
experiments on short carbon nanotube quantum dots.

We thank A. Levy Yeyati for discussions and hospitality during
a visit of F.~S.~at the Universidad Autonoma de Madrid.
This work was supported by the EU.

\end{document}